# Size invariance sector for an agent-based innovation diffusion model

Carlos E. Laciana, Gustavo Pereyra,  Santiago L. Rovere

Grupo de Aplicaciones de Modelos de Agentes (GAMA),
Facultad de Ingeniería, Universidad de Buenos Aires
Avenida Las Heras 2214, Ciudad Autónoma de
Buenos Aires C1127AAR, Argentina.
clacian@fi.uba.ar

**Abstract-** It is shown that under certain conditions it is possible to model a complex system in a way that leads to results that do not depend on system size. As an example of complex system an innovation diffusion model is considered. In that model a set of individuals (the agents), which are interconnected, must decide if adopt or not an innovation. The agents are connected in a member of the networks family known as small worlds networks (SWN). It is found that for a subfamily of the SWN the saturation time and the form of the adoption curve are invariants respect to the change in the size of the system.

## 1. Introduction

Agent-based modeling (ABM) is a relatively new methodology, which is suitable for describing complex adaptive systems [1]. This approach involves a collection of autonomous decision-making entities (agents), an environment through which agents interact, rules that define the relationship between agents and their environment, and rules that determine sequencing of actions in the model [2]. Such agents may be; individuals or companies, in a social context,  which take economic or political decisions  [3-6], animals in ecosystems [7, 8], such as predator-prey systems.

An important use of ABM were  in social sciences, see for example references  [9-11] and more recently in management and organizational behavior research [12-14]. Due to the ductility, the methodology, it results also useful to perform biological models and also for the research in the medicine field. So the agents can be biological constituents as neurons [15], as tumor cells [16, 17],  or being atoms that clump together to form molecules [18].

This wide variety of applications make us considered, agent-based methodology as a new paradigm for simulating natural and social processes [19]. Hence Axelrod [4] calls it "a third way to do science" (in addition to the empirical-experimental methodology and the speculative theoretical work) because it would be like a "theoretical experimentation".

One of the advantages of ABM, respect to other approaches, is the capability of retaining collective behavior which is cause of emerging patterns, such as, for instance, the collective adoption of a political position or the purchase of some product. This methodology lets us analyze which are the parameters that modify the outputs. However there are still some fundamental theoretical aspect that should be taken into account before generalizing the results, as detailed in the reference [20]. For example: the dependence of results with the number of agents ($N$), the extent of the interaction (or the kind of neighborhood of the agents in a network), and the calibration of the time scale.

Regarding the problem of dependence on the number of agents, we will describe, in general, systems with a large number of interacting elements: populations of millions of people, billions of cells, etc. Both from a conceptual and a practical point of view it is interesting to understand how the emerging properties change, in the collective system to be studied, with the changes in the scale of the model.

The type of neighborhood has to do with the configuration of neighbors considered. For instance, for a regular network there might be a von Neumann neighborhood (with 4 connections to first neighbors), Moore (8) or hexagonal (6). This number of connections per agent is what is known as grade and can be thought of as an adjustment variable depending on the problem to be taken into account. Actually, the average degree will be more meaningful, since in many problems (e.g. social modeling) the networks will not be regular. Another feature mentioned is the topology, which depends on the problem at hand.

We have the family of networks known as "small worlds" introduced by Watts and Strogatz [21] examples of which are observed in the neural network of worm Caenorhabditis Elegans, in the USA electric power network, and the collaboration graph of actors in feature films. Other network topologies may be those known as scale-free [22], which are observed in the "World Wide Web" (WWW), and also the block-models stochastic, more elaborate and realistic in social systems [23].

As for setting the time scale, the dynamics of the problem in question will determine whether it is hours, days, months or years. In this sense we can consider it as one more parameter to be adjusted. That is not possible with the number of agents since it is not a parameter that can be adjusted: it is something determined by the actual system to consider. In many cases the complication is that it can be a very large number from a computational standpoint. This is one practical reason to assess the property of invariance respect to the change of system size. That would allow us resolve an analogous problem with a smaller number of agents.

Then the focus of our work is to explore the parameter value space and determine ranges in which we can verify invariance of modeling results against changes of system size. Also we will see if that invariance (against change in the number of agents ($N$) or size) is maintained through changes on the network topology. In the present work, the family of networks known as Small Worlds Networks (SWN) is used, which is characterized by the quantity called rewiring probability ($Pr$). All the family of SWN is obtained, starting from a regular network, by the rewiring, with probability $Pr$, of the connection between agents.

A study concerning to the effects on the dynamics that produces the change in the size of the system is performed in ref. [24]. That study considers a model of opinion formation developed by Galam [25]. Functional dependence of the critical parameters and *N* was found. However, the effect of the network topology was not studied.

In our case we consider a model of diffusion of innovations known as "logit model" [26], inspired by the statistical Ising model [27] which (despite being developed originally in physics) has been used with success in diffusion of innovations [28, 29], with s-type curves as result, which approximate empirical data [30].

The model considered by us can be thought of as a representative model of the cellular automata models [31], which is characterized by a decision algorithm based in threshold. This algorithm is linear in the difference between adopters and not-adopters, coming from the group of relationships of the agent.

The paper is organized as follows: Section 2 provides a brief description of the innovation diffusion model used. Section 3 describes numerical experiments performed. Section 4 summarizes the main conclusions and are discussed the importance of the results obtained and possible future work. There are also two appendixes where indicators are used to quantify the approximations are introduced.

**2. Definition of the model**

In the model under consideration there are two quantities that influence the decision maker; one is due to the influence of the contacts of the agent, by imitation effect, and the other is the evaluation by the agent himself about the benefits of adopting the innovation. This behavior can be summarized in the following mathematical algorithm: the agent *i* adopts the innovation if $\Delta_i > 0$, with $\Delta_i = \Delta v_i + \Delta u_i$, where $\Delta v_i$ is the proportion of neighbors (or contacts) of agent *i* that adopt the innovation minus the proportion of those who do not. The term $\Delta u_i$ is the utility difference between adopting and not adopting perceived by agent i. This model is analogous to the statistical model known as Ising model at zero temperature, with an extern field. It was originally developed to study phase transitions in ferromagnetic materials [27]. In the analogy the metal atoms correspond to the decision makers, who can adopt (spin up) or not adopt (spin down) the innovation.

In this model the agents are connected by a two-dimensional Small-World Network [32]. That network is built using the Watts-Strogatz procedure [21] from a regular two-dimensional network with periodic boundary conditions (toroidal), for a set of values of the rewiring probability $P_r$. The periodic boundary conditions let us avoid differences in the possibility of reaching the threshold between the agents which are on the edge and those in the interior of the network. The re-wiring takes place in the initial preparation of the network, making (with probability $P_r$) each agent break a connection with one of its neighbors in the regular network, to connect with another agent at random. In the process the mean degree does not change.

## 3. Results

The numerical tests have been made keeping the same proportion of initial adopters (or early adopters) for different grid sizes. The values considered were 2.5% and 10%. The first value is inspired on the theory of Rogers [33] on diffusion of innovations and the second is considered to analyze scale invariance against changes in the size of the initial seed. The seed of adopters is introduced before the start of the simulation, i.e. preparing the system with initials adopters randomly and uniformly distributed. The value of the difference of utility was chosen arbitrarily equal to 0.6 for all the agents (homogeneous population).

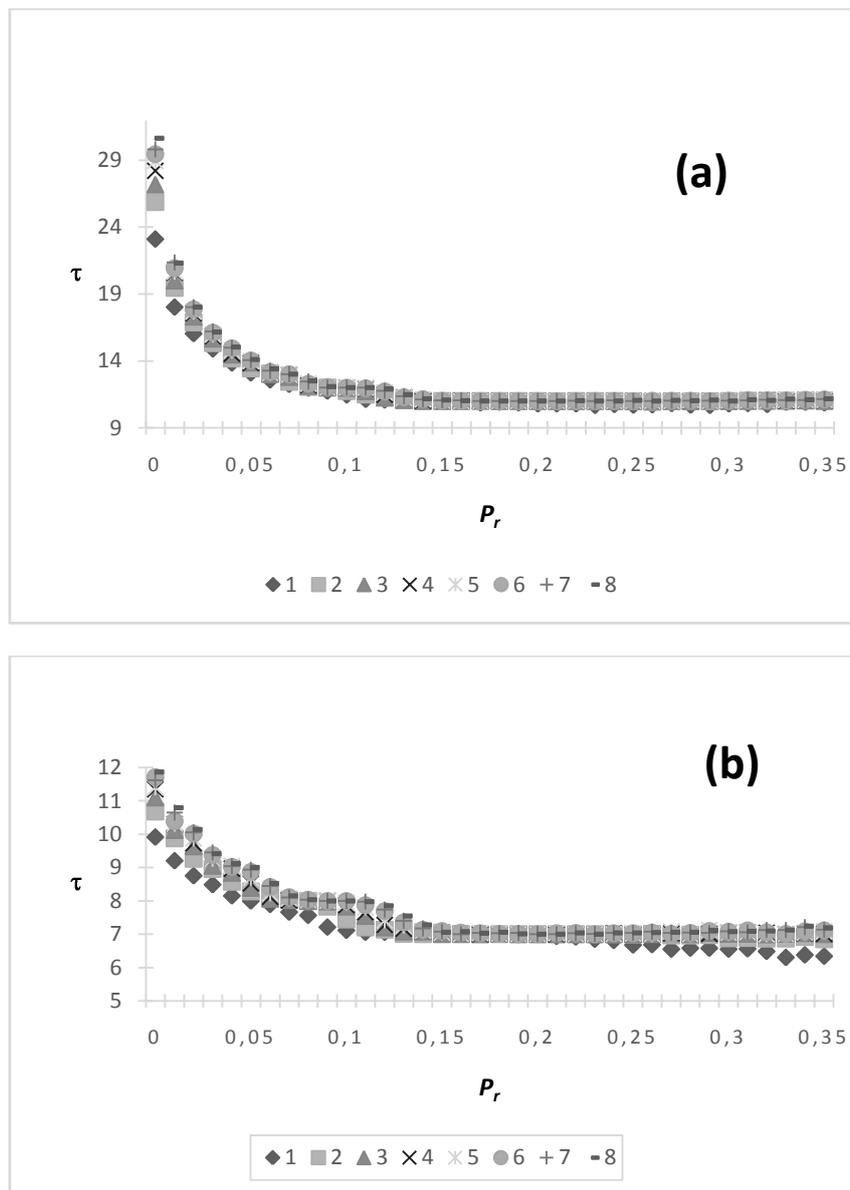

*Figure 1.* Simulation results for the Ising-like model for Innovation Diffusion over a range of Small-World non-directed networks constructed using the Watts-Strogatz method, with periodic boundary conditions (toroidal), Moore neighborhood (grade 8), and different population sizes (N) from $100^2$ up to $800^2$. Graphs in Figures 1a and 1b show the Saturation Time ($\tau$) variation across different network layouts, represented by the values of the Rewiring Probability ($P_r$) used to construct each network. Each population size (N) is labeled with $\sqrt{N}/100$. Series averaging 50 simulation runs of the model using REPAST Simphony, an open source framework maintained by Argonne

In Fig. 1, we can see, the average curves of saturation time ($\tau$) vs. rewiring probability $P_r$ for the interval [0, 0.35]. Each curve corresponds to a grid size $N$, with values $100^2$, $200^2$,...., $800^2$. In Fig. 1a (corresponding to 2.5% of early adopters) the important result obtained is that from $Pr \geq 0.14$ the shape of the curve does not depend on $N$, we can vary the size of the system without producing a change in the time of adoption. In Fig. 1b (corresponding to 10% of early adopters) the interval of invariance obtained (as we can see) is $Pr \in [0.15, 0.21]$.

In order to give more precision, about the difference, between the curves $\tau$ vs $P_r$ shown in Fig. 1a,b we introduce a quantity which is coming from the average of the differences, in the following form:

$$\sigma_\tau(P_r) = \frac{1}{S}\left\{[\tau_1(P_r) - \tau_S(P_r)]^2 + \sum_{l=2}^{S}[\tau_l(P_r) - \tau_{l-1}(P_r)]^2\right\}^{1/2} \quad (1)$$

Where $\tau_l(P_r)$ is the saturation time, for the size of network indicated by "$l$", i.e. $N_l = N_1, N_2,...., N_S$ (with $S$ the total number of samples, in our case $S=8$). Using the last equation we obtain the curves of Fig. 2a, b:

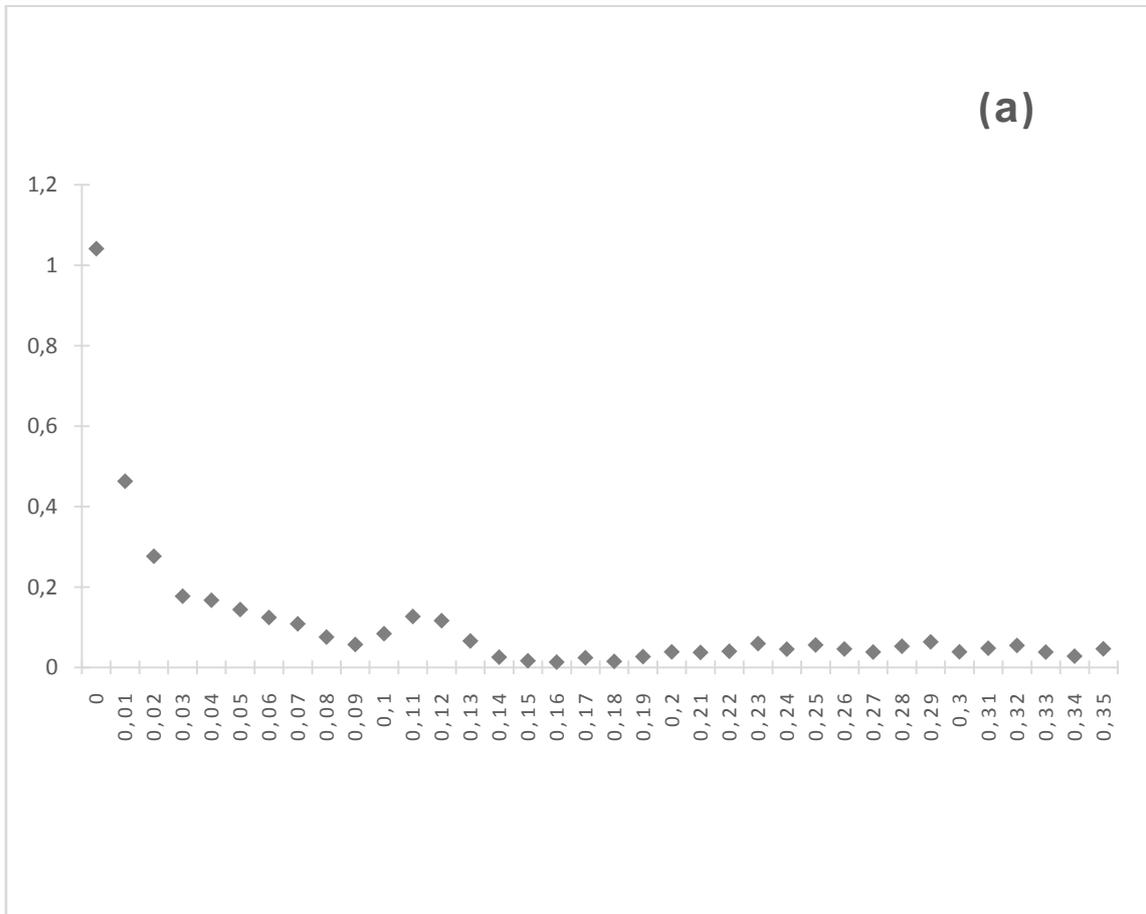

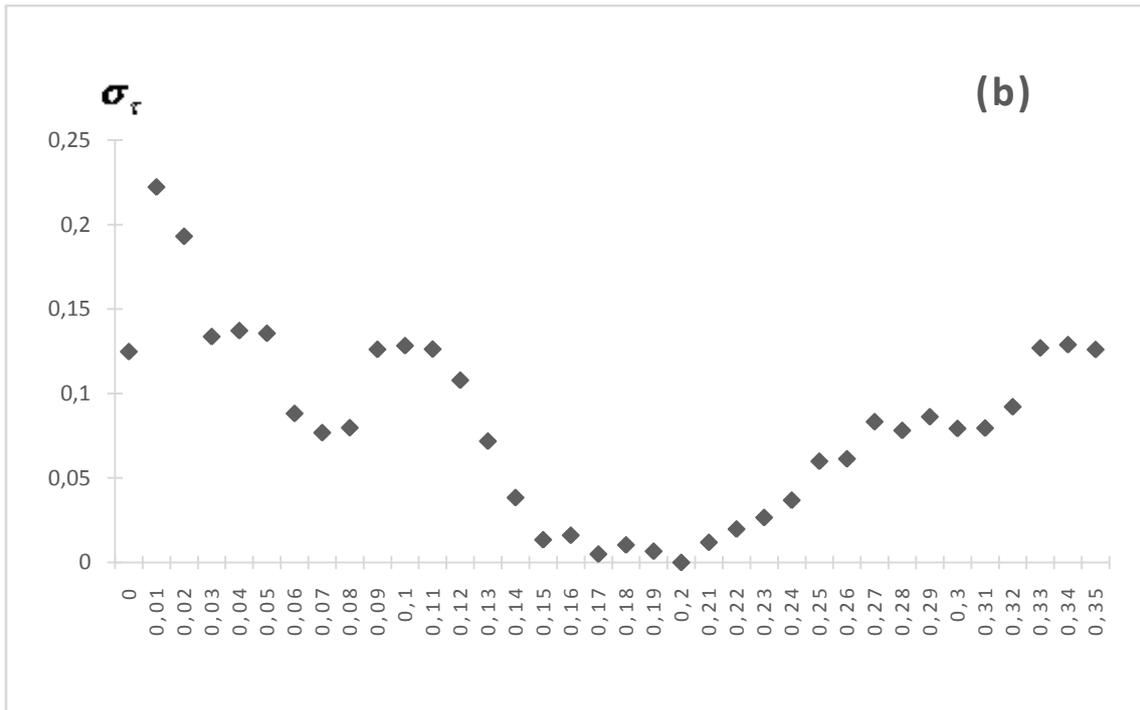

*Figure 2a, b*: In the figure are represented $\sigma_\tau$ vs $P_r$, given by Eq. (1), for 2.5% and 10% of early adopters respectively.

To show that not only is valid the coincidence observed, for the saturation time, in Fig. 1 (between curves with different N), in Fig. 3 it is shown also the coincidence of the diffusion curves, for a particular value of $P_r$ from the interval of invariance.

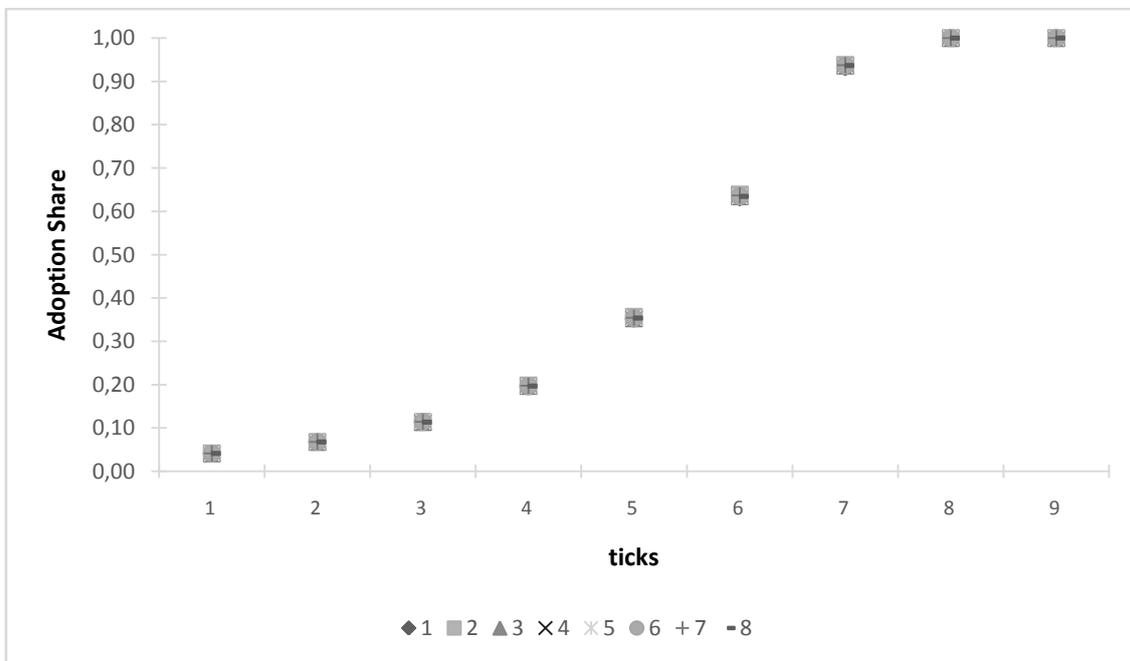

*Figure 3:* Innovation diffusion (Adoption Share) over time (ticks) averaging 50 simulation runs for Small-World networks with Pr=0.18. As in the previous figures, each series corresponds to a different population size (N) labeled with $\sqrt{N}/100$.

Figure 3 shows how the overlap between diffusion curves for different values of *N* is such that it is impossible to distinguish between different system sizes. For that reason, in appendix B is introduced an indicator that let us appreciate the degree of coincidence of the results. This shows that complex systems, interconnected by a network of small worlds, with $P_r$ in a range of, approximately, 0.15 - 0.21 (including the example considered in detail) can be simulated with $N = 100^2$, still if in the real problem the size is more bigger.

## 4. Conclusions

We have seen that the invariance of the saturation time to changes in system size, is sensitive to changes in the probability of re-wiring of the SWN, where agents interact, and to the size of the seed of the early adopters.

More precisely when $P_r \geq 0.14$ and the seed of early adopters is 2.5% and when $P_r \in [0.15, 0.21]$ and the percentage of early adopters is 10%, saturation times, for different network sizes, coincide with an error less than 6%. This behavior can relate to a characteristic parameter of the network which is known as clustering coefficient ($C$), it suffices to look at the curve given in ref. [34] which relates $C$ with the probability of rewiring, then we see that for the first case of study corresponds to $C \leq 0,7$, while the second is approximately between 0.5 and 0.6. The first case, as shown in the table of the ref. [35] describes the WWW networks and many social networks of scientific collaboration. In the second case reaches only networks of scientific collaboration.

Clearly, the existence of sectors with invariance against resizing, of the set of agents, let us the use smaller systems without loss of physical content and with an economies of computational time in the modeling.

It would be interesting, as future work, extend the study including nonlinear decision models (see ref. [36]) to see if the invariance is an intrinsic property of linearity or is more strongly determined by the topology of the network.

**Appendix A**

Let us give a criterion to decide how many saturation time curves which are necessary to get an average curve representative of the diffusion process. In order to do that we choose the values of $P_r$ = 0, 0.05 and 0.35 (low, medium and high) in the curves corresponding to different averages. Then we consider the saturation time as a function of two variables, i.e. $\tau = \tau(\Gamma, P_r)$, where $\Gamma$ is the number of curves uses to perform the average. Moreover we define: $\tau_1(\Gamma) \equiv \tau(\Gamma, P_r = 0)$, $\tau_2(\Gamma) \equiv \tau(\Gamma, P_r = 0.05)$, $\tau_3(\Gamma) \equiv \tau(\Gamma, P_r = 0.35)$ and also an indicator ($\Delta\tau(\Gamma)$) of the difference between saturation times, coming from the average performed using the number of curves $\Gamma$ and $\Gamma + 1$, i.e.:

$$\Delta\tau(\Gamma) = [(\tau_1(\Gamma+1) - \tau_1(\Gamma))^2 + (\tau_2(\Gamma+1) - \tau_2(\Gamma))^2 + (\tau_3(\Gamma+1) - \tau_3(\Gamma))^2]^{1/2}$$

.

In the following figure we can see Δτ vs Γ:

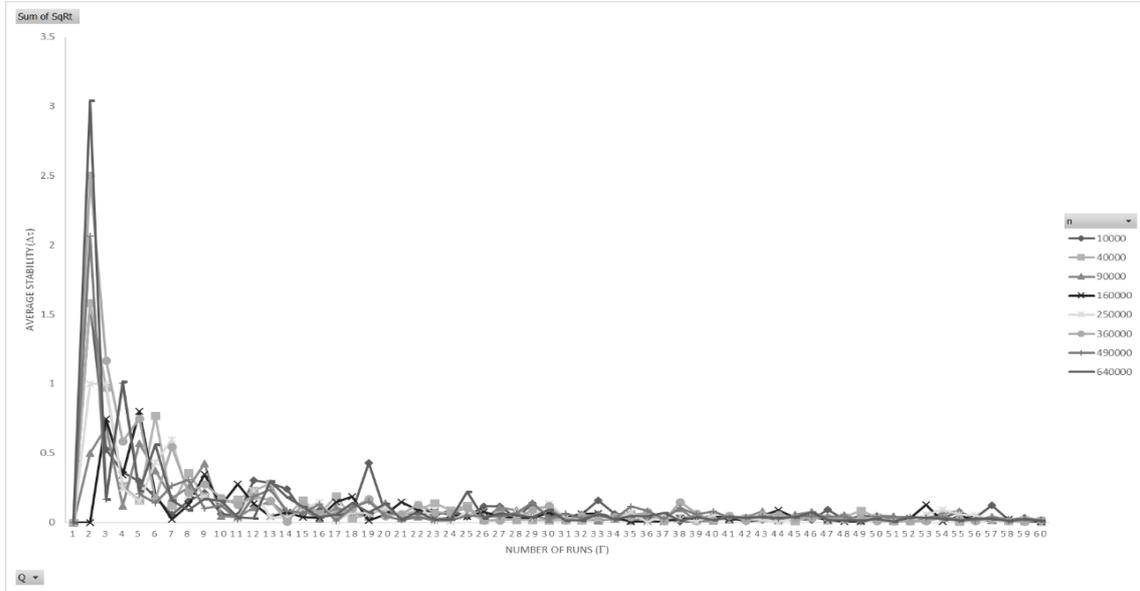

*Figure 4:* Average Stability (Δτ) vs Number of Runs (Γ). A fast convergence is observed, for size networks of 10000, 40000, 90000, 160000, 250000, 360000, 490000, and 640000 agents. Only is considered the case with a seed of 2.5% of early adopters.

We can see, in the last figure, a fast convergence. This ensures that the sample of Γ = 50 used in our calculations is a good approximation.

**Appendix B**

We define the indicator $\sigma_n(t)$ in order to compare the diffusion curves (*n* vs *t*) obtained for different sizes of network and for the particular value $P_r = 0.18$, as benchmark value. The functional form is:

$$\sigma_n(t) = \frac{1}{S}\left\{[n_1(t) - n_S(t)]^2 + \sum_{l=2}^{S}[n_l(t) - n_{l-1}(t)]^2\right\}^{1/2},$$

where $n_l(t) \equiv n(t,l)$, with $l \equiv \sqrt{N}/100$ and $S = 8$, which is the label of the network size.
In the following figure we can see the curve $\sigma_n$ vs t:

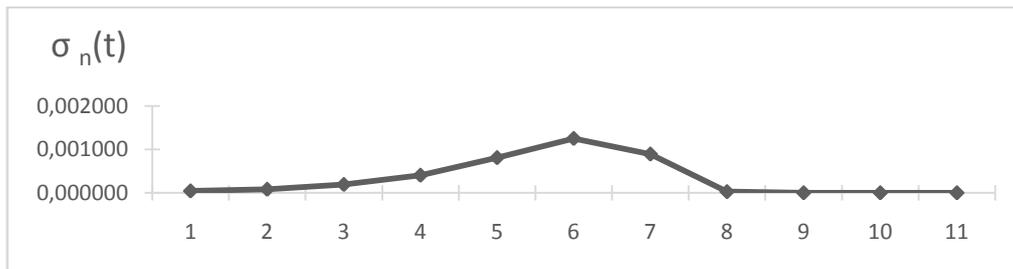

*Figure 5:* Deviation indicator for the calculation of diffusion curve due to the change in network size.

As we can see in the figure the bigger $\sigma_n$ is approximately 0.0012. That separation is produced at 6 ticks, which, in the diffusion curve (Fig. 3), corresponds to the maximum slope.

**Acknowledgements**

This research was supported partially by two US National Science Foundation (NSF) Coupled Natural and Human Systems grants (0410346 and 0709681) and by the University of Buenos Aires grant: UBACyT-01/Q710.